\documentclass[a4paper]{article}
\RequirePackage[english]{babel}
\RequirePackage[latin1]{inputenc}
\RequirePackage[T1]{fontenc}
\RequirePackage{mathrsfs}
\RequirePackage{amsmath}
\RequirePackage{amssymb}
\RequirePackage{amsbsy}
\RequirePackage{bm}
\begin{document}
\title{\bf{On the Principle of Equivalence}}
\author{Luca Fabbri\\ 
\footnotesize Groupe Physique des Particules, D\'{e}partement de Physique de l'Universit\'{e} de Montr\'{e}al\\
\footnotesize Gruppo Teorici, Dipartimento di Fisica dell'Universit\`{a} di Bologna \& INFN}
\date{}
\maketitle
\ \ \ \ \textbf{PACS}: 04.20.Cv (Fundamental problems and general formalism)
\begin{abstract}
We consider the Principle of Equivalence along with Weyl theorem to discuss the interpretation of gravity as a geometric effect; we study what are the restrictions on the connections that must be required for this geometrization to occur in the most general case.
\end{abstract}
\section*{Introduction}
The Principle of Equivalence states that any gravitational field can be compensated by local inertial forces, so that the gravitational-inertial effects can be vanished in a given spacetime point of a given reference system; on the other hand differential geometries are endowed with peculiar entities called connections that can be made to vanish in a given spacetime point of a given reference system: this makes clear that the connections and the gravitational field behave in similar ways. However for the removal of any gravitational field it is reasonable to suppose that there always exists a unique choice of a given spacetime point in a given reference system, whereas for general connections this choice may not be unique nor exist at all: general connections and the gravitational field do not behave in exactly the same way. Thus restrictions must be imposed upon general connections in order to reduce them to a form for which such a choice is unambiguously defined, so to have these special connections and the gravitational field behave in exactly the same way. This would allow the representation of the gravitational field within those special connections, giving rise to the interpretation of gravity as a geometric effect.

On the other hand, for the connections we have a result known as Weyl theorem stating that no matter how general a connection is, nevertheless it is always possible for its symmetric parts to be removed in a given point of a given reference system. Therefore it is Weyl theorem that suggests what restrictions upon general connections have to be imposed.

Obviously by restricting the connection to be metric symmetric we would get the connection with only one symmetric part solving the problem straightforwardly, and in fact in most of the commonly used textbooks metric symmetric connections are assumed, although these constraints are assumed more as a way to simplify the structure of the geometry than as fundamental assumptions; thus although the above solution is indeed a solution, nevertheless such solution is too restrictive, and some assumptions should be loosened to let more general connections be defined.

Indeed the requirement for which the connections have to be metric symmetric is not a fundamental assumption at all, and in fact part of the literature does not assume it in order to deal with more general connections; however on the other hand exaggerated generalizations of the connections have the consequence of recreating the situation in which too general connections and the gravitational field are too different to permit the geometrization of gravity.

So restrictions have to be imposed and they cannot be too strong, so to have the most general theory in which gravity is geometrized.

In the present paper, we will consider the problem of finding such most general restrictions.
\section{Geometric Restrictions from\\ the Principle of Equivalence}
The Principle of Equivalence states that any given gravitational field has effects that can be compensated by local inertial effects, so that the gravitational-inertial effects can be vanished in a spacetime point by a suitable choice of the coordinate system; and further, the widely known Weyl theorem states that general connections have symmetric parts that yield a null result when calculated in a spacetime point of a given coordinate system: this means that the symmetric parts of the connections is where the gravitational information has to be contained. However, while the gravitational field should physically be uniquely determined, the symmetric part of the connection is not: thus the problem is that we do not know in which one of the symmetric parts of the connection the gravitational information is contained. Hence special restrictions must be required for the symmetric parts of the connection in order for them to be uniquely defined, so that the only symmetric part of the connection will necessarily be where the gravitational information is stored. This would induce the representation of the gravitational field within the metric symmetric connection, giving rise to the interpretation of gravity as a geometric effect.

These ideas need now be expressed in more rigorous terms, and to this purpose we shall introduce the foundations of the differential geometry.

This geometry is built upon two fundamental tensors, the Metric Tensors $g_{\mu\nu}$ and $g^{\mu\nu}$ such that $g_{\mu\nu}=g_{\nu\mu}$ and $g^{\mu\nu}=g^{\nu\mu}$ and also such that they verify $g_{\mu\nu}g^{\nu\rho}=\delta_{\mu}^{\rho}$ where the $\delta_{\mu}^{\rho}$ is the Kronecker Unity Tensor; the enlargement of the geometry to differential geometry is achieved by means of the Connection $\Gamma^{\alpha}_{\mu\nu}$ defined in terms of its transformation law, so that after its definition the connection is used to construct the Covariant Derivative $D_{\mu}$ acting on tensors and yielding tensors: with these fundamental quantities we define
\begin{equation}
\Lambda^{\rho}_{\mu\nu}=\frac{1}{2}g^{\rho\sigma}
(\partial_{\nu}g_{\mu\sigma}+\partial_{\mu}g_{\nu\sigma}-\partial_{\sigma}g_{\mu\nu})
\end{equation}
which is a connection called Levi-Civita Connection, then we define the tensor
\begin{equation}
L^{\rho}_{\mu\nu}=
\frac{1}{2}g^{\rho\sigma}(D_{\nu}g_{\mu\sigma}+D_{\mu}g_{\nu\sigma}-D_{\sigma}g_{\mu\nu})
\end{equation}
and by using the connection alone it is possible to define the entity given by $\Gamma^{\rho}_{\alpha\beta}-\Gamma^{\rho}_{\beta\alpha}=Q^{\rho}_{\phantom{\rho}\alpha\beta}$ which is a tensor called Cartan Torsion Tensor and with which we finally define the tensor
\begin{equation}
K^{\rho}_{\mu\nu}=
\frac{1}{2}\left(Q_{\mu\nu}^{\phantom{\mu\nu}\rho}+Q_{\nu\mu}^{\phantom{\nu\mu}\rho}\right)
\end{equation}
so that with all these quantities we have
\begin{equation}
\Gamma^{\rho}_{\mu\nu}=\Lambda^{\rho}_{\mu\nu}
-L^{\rho}_{\mu\nu}+K^{\rho}_{\mu\nu}
+\frac{1}{2}Q^{\rho}_{\phantom{\rho}\mu\nu}
\label{dec}
\end{equation}
as the decomposition of the most general connection possible. The symmetry properties in the two lower indices and the transformation law defining the connection have an interesting relationship expressed by Weyl Theorem, which states that it is possible to find a suitable choice of coordinates in which locally the symmetric part in the two lower indices of a general connection is equal to zero: however, from (\ref{dec}) we see that the symmetric parts are given by the largest $\Lambda^{\rho}_{\mu\nu}-L^{\rho}_{\mu\nu}+K^{\rho}_{\mu\nu}$ but also by smaller parts such as $\Lambda^{\rho}_{\mu\nu}-L^{\rho}_{\mu\nu}$ and $\Lambda^{\rho}_{\mu\nu}+K^{\rho}_{\mu\nu}$ and the smallest Levi-Civita connection $\Lambda^{\rho}_{\mu\nu}$ as expected, and for each and every one of them Weyl theorem furnishes the procedure to choose coordinates in which locally they are equal to zero. Thus to make the symmetric parts of the connection uniquely defined we have to make them all collapse onto one symmetric part alone by demanding the restrictions given by the conditions $L^{\rho}_{\mu\nu}=0$ and $K^{\rho}_{\mu\nu}=0$ or equivalently $D_{\rho}g_{\mu\nu}=0$ and $Q_{\rho\mu\nu}=Q_{[\rho\mu\nu]}$ so that the metric symmetric Levi-Civita connection $\Lambda^{\rho}_{\mu\nu}$ is the only symmetric part of the connection, the one in which the gravitational degrees of freedom are then found necessarily. This induces the representation of the gravitational degrees of freedom within the Levi-Civita connection and thus within the metric tensor, giving rise to the interpretation of gravity in terms of metric effects.

Thus said we have that the connection given by
\begin{equation}
\Gamma^{\rho}_{\mu\nu}=\frac{1}{2}g^{\rho\sigma}
(\partial_{\nu}g_{\mu\sigma}+\partial_{\mu}g_{\nu\sigma}-\partial_{\sigma}g_{\mu\nu})
+\frac{1}{2}Q^{\rho}_{\phantom{\rho}\mu\nu},\ \ \ \ \ \ \ \ Q_{\rho\mu\nu}=Q_{[\rho\mu\nu]}
\label{conn}
\end{equation}
is such that its only symmetric part is written in terms of the metric tensor and so the interpretation of gravity within the metric tensor is necessary, and such interpretation is based on arguments confined into the metric symmetric part of the connection alone and therefore it is not affected by the presence of a completely antisymmetric Cartan torsion tensor; by converse the addition of a completely antisymmetric Cartan torsion tensor does make this connections the most general possible.

We remark that our result has been obtained by using arguments on the connection of the spacetime alone, and so they are valid even if the connection of the spacetime is further generalized to the connection of a complex representation of the spacetime, eventually after the inclusion of the electrodynamic gauge potential, as spinorial connection.

Therefore we have that these connections are restricted enough the make the interpretation of gravity within the metric tensor necessary and among all such connections they are the most general possible.

To compare this result with others, we notice that the condition of complete antisymmetry of Cartan torsion can be obtained by postulating the condition of metricity, as discussed in \cite{f}; but by its side, the condition of metricity is well-known to be related to the Minkowskian structure of the spacetime and its Lorentz invariance under transformations of coordinate systems, and then to the Principle of Equivalence and Causality, as discussed by Hayashi in \cite{hay} and by Hehl, Von Der Heyde, Kerlick and Nester in \cite{h-h-k-n}: so it is not surprising that both the completely antisymmetric Cartan torsion and metricity can be obtained at once from the Principle of Equivalence and Causality, as it has also been discussed by Macias and L\"{a}mmerzahl in \cite{m-l}. Also we notice that in \cite{xy} Xin Yu proves the same result using the supplementary hypothesis for which all the different symmetric parts of the connection have to be zero in the same point of a given coordinate system, but this result is weaker than ours because the supplementary hypothesis is not necessarily ensured.

Now if we consider our result trying to restrict it even further we have that we could only require Cartan torsion tensor to be zero.

Some authors have tried to show that the vanishing of torsion is linked to physical principles, and in order to follow their reasoning we recall that we can shift the discussion from the gravitational field of acceleration to the gravitational acceleration felt by material bodies. The law that is used to describe the motion of material bodies is Newton's Law, whose covariant form is given upon defining the infinitesimal displacement along the coordinate axes $dx^{\mu}$, thus the infinitesimal displacement along a general direction called line element and indicated with $ds^{2}=g_{\mu\nu}dx^{\mu}dx^{\nu}$, and finally the covariant expression of the velocity vector given by $u^{\mu}=\frac{dx^{\mu}}{ds}$, so that Newton's Law in covariant form is given by
\begin{equation}
u^{\alpha}D_{\alpha}u^{\mu}=0
\end{equation}
in the free case: its non-covariant expansion in terms of the non-covariant acceleration $w^{\mu}=\frac{du^{\mu}}{ds}$ is given by
\begin{equation}
w^{\mu}+\Gamma^{\mu}_{\rho\alpha}u^{\rho}u^{\alpha}=0
\end{equation}
called Autoparallel Equation, describing the straightest trajectory between two given points, and in which it is clear that in the spurious term because of the presence of the symmetric tensor $u^{\rho}u^{\alpha}$ only the symmetric part of the connection is selected; this equation reduces to
\begin{equation}
w^{\mu}+\Lambda^{\mu}_{\rho\alpha}u^{\rho}u^{\alpha}=0
\end{equation}
called Geodesic Equation, describing the shortest trajectory between two given points, as in the torsionless case. So the motion of material bodies governed by Newton's Law is not affected by a completely antisymmetric torsion: as Newton's Law governs the dynamics of material bodies in macroscopic situations then in macroscopic situations torsional degrees of freedom are lost. It is interesting to consider that in \cite{we} Weinberg starts from Newton's Law to show that torsion vanishes in the most general theory of gravitation; this could make one believe that torsion is equal to zero, but careful analyses show that it is precisely because it is based on Newton's Law that torsion vanishes in this approach, and since Newton's Law is substantially macroscopic then we conclude that in this book the author proves that in macroscopic situations torsion is equal to zero, which is what is to be expected. In \cite{we} Weinberg follows the approach motivated by having a theory of gravity that may be compatible with principles of the microscopic domain whereas it is in terms of macroscopic physics that his discussion is entirely outlined, which is the reason for which also in other textbooks like \cite{m-t-w} and \cite{wa} Misner, Thorne and Wheeler and Wald still consider torsion to be absent, that is torsion is not needed precisely because physical laws of macroscopic validity alone are considered, which is expected and well-known. Thus the vanishing of torsion cannot be obtained when also microscopic domains are considered in the physical picture.

In this way we have shown that torsion is present in the most general situation, and if we consider our result trying to extend it even further we would lose the interpretation of gravity as geometric effect.

Indeed extensions of our result would be unable to implement the Principle of Equivalence via Weyl theorem. This would produce inconsistencies in situations in which the interpretation of gravity as geometry is clearly valid, as explained in \cite{a-p-v}, \cite{a-b-p} and \cite{a-p} by Aldrovandi, Pereira, Barros and Vu.

So torsion is present in the most general situation and it has to be completely antisymmetric and within metric connections to give rise to the interpretation of gravity as geometric effect compatibly with all observations.
\section{Physical Constraints of\\ the Principle of Equivalence}
So far we have discussed how torsion has to be included for generality and it has to be completely antisymmetric within metric connections in order to be consistent with the interpretation of gravity in terms of geometry compatibly with the limits imposed by the present observations.

In order to consider this geometry as the background for an eventual physical theory, we will follow the same spirit that led Einstein to consider the purely metric geometry as the background for the eventual gravitational theory and simply extend it to the inclusion of torsion; then we will consider torsion to be completely antisymmetric: after this generalization we will have that the Einstein tensor $E^{\mu\nu}$ will have an antisymmetric part written in terms of the Cartan torsion $Q^{\rho\mu\nu}$ and both these tensors will satisfy the geometric identities of Jacobi-Bianchi, so that in terms of the energy $T^{\sigma\rho}$ and the spin $S^{\nu\sigma\rho}$ and with coupling constant $k$ the system
\begin{eqnarray}
E^{\sigma\rho}=-\frac{k}{2}T^{\sigma\rho}
\label{einstein}
\end{eqnarray}
and
\begin{eqnarray}
Q^{\nu\sigma\rho}=kS^{\nu\sigma\rho}
\label{sciama-kibble}
\end{eqnarray}
is postulated to be the system of field equations; as we have completely antisymmetric torsion then we will have completely antisymmetric spin so that the spin $S^{\nu\sigma\rho}$ will be such that
\begin{eqnarray}
S^{\nu\sigma\rho}=S^{[\nu\sigma\rho]}
\label{constraint}
\end{eqnarray}
for the system of field equations of gravitational spin-constrained spinorial field theories. This is the theory that we have to find eventually, but before proceeding there is an important point that needs to be addressed now.

This issue concerns the introduction of electrodynamics as a gauge theory based on the definition of the Maxwell tensor
\begin{eqnarray}
F_{\mu\nu}=\partial_{\mu}A_{\nu}-\partial_{\nu}A_{\mu}
\label{Maxwell}
\end{eqnarray}
invariant under the gauge transformation 
\begin{eqnarray}
A'_{\nu}=A_{\nu}-\partial_{\nu}\phi
\label{gauge}
\end{eqnarray}
as it is well-known: if the generalization to the torsional case were given by having a Maxwell tensor defined as the antisymmetric part of the covariant derivative then we would have
\begin{eqnarray}
\Phi_{\mu\nu}=D_{\mu}A_{\nu}-D_{\nu}A_{\mu}
\equiv \partial_{\mu}A_{\nu}-\partial_{\nu}A_{\mu}+A_{\rho}Q^{\rho}_{\phantom{\rho}\mu\nu}
=F_{\mu\nu}+A_{\rho}Q^{\rho}_{\phantom{\rho}\mu\nu}
\label{Maxwellgeneralized}
\end{eqnarray}
which would not be invariant for the gauge transformation above but for the generalized gauge transformation
\begin{eqnarray}
A'_{\nu}=A_{\nu}-e^{\phi}\partial_{\nu}\phi
\label{gaugegeneralized}
\end{eqnarray}
although such a generalization would also require torsion to be given by
\begin{eqnarray}
Q^{\alpha}_{\phantom{\alpha}\mu\nu}=\frac{1}{3}
\left(\delta^{\alpha}_{\mu}\partial_{\nu}\phi-\delta^{\alpha}_{\nu}\partial_{\mu}\phi\right)
\end{eqnarray}
which is not completely antisymmetric; on the other hand that the gauge theory should not be generalized as to have the Maxwell tensor defined as the antisymmetric part of the covariant derivative is clear from the fact that the Maxwell tensor must be defined as the commutator 
\begin{eqnarray}
[D_{\mu},D_{\nu}]\psi=iF_{\mu\nu}\psi
\end{eqnarray}
of the gauge covariant derivatives
\begin{eqnarray}
D_{\mu}\psi=\partial_{\mu}\psi+iA_{\mu}\psi
\end{eqnarray}
and in this way we see that the definition of the Maxwell tensor is the one we have in the torsionless case
\begin{eqnarray}
F_{\mu\nu}=\partial_{\mu}A_{\nu}-\partial_{\nu}A_{\mu}
\end{eqnarray}
invariant under the gauge transformation 
\begin{eqnarray}
A'_{\nu}=A_{\nu}-\partial_{\nu}\phi
\end{eqnarray}
and in which the decoupling between electrodynamics and torsion is clearly a property of the system in the most general situation. With this construction it is now easy to see what geometric identities are obtained, and in what way they shall be used in the following in order for the physical theory to be postulated.

Indeed this form of the Maxwell tensor is such that it satisfies the geometric identity given by
\begin{eqnarray}
D_{\rho}\left(D_{\sigma}F^{\sigma\rho}+\frac{1}{2}F_{\alpha\mu}Q^{\alpha\mu\rho}\right)=0
\label{gaugefield}
\end{eqnarray}
so that in terms of the current $J^{\rho}$ along with the energy $T^{\sigma\rho}$ and spin $S^{\nu\sigma\rho}$ and with coupling constant $k$ the generalized system
\begin{eqnarray}
D_{\sigma}F^{\sigma\rho}+\frac{1}{2}F_{\alpha\mu}Q^{\alpha\mu\rho}=J^{\rho}
\label{electrodynamics}
\end{eqnarray}
along with
\begin{eqnarray}
E^{\sigma\rho}
+\frac{k}{2}\left(\frac{1}{4}g^{\sigma\rho}F^{2}-F^{\sigma\mu}F^{\rho}_{\phantom{\rho}\mu}\right)
=-\frac{k}{2}T^{\sigma\rho}
\label{gravity}
\end{eqnarray}
and
\begin{eqnarray}
Q^{\nu\sigma\rho}=kS^{\nu\sigma\rho}
\label{spinor}
\end{eqnarray}
such that
\begin{eqnarray}
S^{\nu\sigma\rho}=S^{[\nu\sigma\rho]}
\label{constraints}
\end{eqnarray}
is postulated to be the system of field equations for gravitational electrodynamic spin-constrained spinorial field theories. This is the theory we have to find now.

To show that such a gravitational electrodynamic spin-constrained spinorial field theory really exists it is enough to mention that actually the case represented by massive charged spin-$\frac{1}{2}$ spinorial field theories is a theory that fits precisely in the scheme described here.
\section*{Conclusion}
In the present paper, we have shown that spacetime connections that are metric with completely antisymmetric Cartan torsion tensors are connections with a unique symmetric part written in terms of the metric tensor, so that the interpretation of gravity as a metric field is necessary and among all the connections for which gravitation is geometrized these connections are the most general possible; by employing such connections, the Principle of Equivalence is obtained as the physical translation of what Weyl theorem tells about the geometry: further it has been discussed that using different connections would result in two complementary diseases, either in a loss of generality, or in the exclusion of the Principle of Equivalence from the geometry, and while the former is philosophically unacceptable, the latter is ruled out by observations. Finally we have shown that such connections define a geometric relativistic theory upon which it is possible to construct physical theories, by finding for instance that fermions have place in this scheme, and intriguingly this is the one and only matter theory that has been directly observed in nature.

\

\noindent \textbf{Acknowledgments.} I wish to thank Professor Manu B. Paranjape and Professor Richard MacKenzie for their constructive critics, and University of Montr\'{e}al for kind hospitality; this work has been financially supported by NSERC of Canada.

\


\end{document}